# Economic and environmental impacts of ballast water management on Small Island Developing States and Least Developed Countries

Zhaojun Wang[1,*], Amanda M. Countryman[2], James J. Corbett[3], Mandana Saebi[4]

[1] *School of Marine Science and Policy, College of Earth, Ocean, and Environment, University of Delaware, Delaware, USA*
[2] *Department of Agricultural and Resource Economics, Colorado State University, Colorado, USA*
[3] *School of Marine Science and Policy, College of Earth, Ocean, and Environment, University of Delaware, Delaware, USA*
[4] *University of Notre Dame, 384 Nieuwland, Notre Dame, IN 46556, USA*

*Corresponding author
Author e-mails: izhaojun@udel.edu (ZW), Amanda.Countryman@colostate.edu (AMC), jcorbett@udel.edu (JJC), Mandana.Saebi.1@nd.edu (MS)*

**Abstract**

  The Ballast Water Management Convention can decrease the introduction risk of harmful aquatic organisms and pathogens, yet the Convention increases shipping costs and causes subsequent economic impacts. This paper examines whether the Convention generates disproportionate invasion risk reduction results and economic impacts on Small Island Developing States (SIDS) and Least Developed Countries (LDCs). Risk reduction is estimated with an invasion risk assessment model based on a higher-order network, and the effects of the regulation on national economies and trade are estimated with an integrated shipping cost and computable general equilibrium modeling framework. Then we use the Lorenz curve to examine if the regulation generates risk or economic inequality among regions. Risk reduction ratios of all regions (except Singapore) are above 99%, which proves the effectiveness of the Convention. The Gini coefficient of 0.66 shows the inequality in risk changes relative to income levels among regions, but risk reductions across all nations vary without particularly high risks for SIDS and LDCs than for large economies. Similarly, we reveal inequality in economic impacts relative to income levels (the Gini coefficient is 0.58), but there is no evidence that SIDS and LDCs are disproportionately impacted compared to more developed regions. Most changes in GDP, real exports, and real imports of studied regions are minor (smaller than 0.1%). However, there are more noteworthy changes for select sectors and trade partners including Togo, Bangladesh, and Dominican Republic, whose exports may decrease for textiles and metal and chemicals. We conclude the Convention decreases biological invasion risk and does not generate disproportionate negative impacts on SIDS and LDCs.

## 1. Introduction

  Global shipping plays an important role in international trade. Shipping activities, however, generate negative environmental impacts on port and coastal ecosystems through the introduction of alien species via ballast water discharge and biofouling (Miller and Ruiz, 2014).

Once nonindigenous species (NIS) become established, they may cause serious harm to human health, ecosystems, and the economy (Carlton, 2003, Ruiz and Carlton, 2003, Keller et al., 2008, Keller et al., 2009). Ballast water discharge is one vector of species introduction via human activities. It is important to know the relation between humans and biological invasions, including how people cause biological invasions and how people respond to them (Shackleton et al., 2019). In the context of ballast water-mediated harmful aquatic organisms and pathogens (HAOP), the 2004 Ballast Water Management (BWM) Convention at the International Maritime Organization (IMO) is applied at the international level. The technological solution to NIS risk reduction is to use a Ballast Water Management System (BWMS) to achieve the required numeric concentration for different organisms.

Studies examine the effects of the BWM Convention from different perspectives, including technological efficacy, cost-effectiveness, and economic impacts. Many studies conduct experiments to measure organism concentrations in treated and untreated ballast water to determine efficacies of different BWMS (Ruiz and Reid, 2007, Briski et al., 2015, Cohen and Dobbs, 2015, Herwig et al., 2006, Reusser et al., 2013, Minton et al., 2005). Some estimate the regulatory compliance costs. For example, King et al. find a preliminary estimate of the vessel-based BWMS using various methods (i.e. mechanical, chemical, and physical methods) (King et al., 2009) and Glosten et al. conduct a study to estimate potential costs of barge-based BWMS designed to meet the stricter ballast water discharge standards of California (Glosten et al., 2018). Wang and Corbett examine the cost-effectiveness of technology to meet different ballast water management policies (Wang and Corbett, 2020). Fernandes et al. estimate costs of ballast water treatment for European shipping and benefits of fuel cost-saving (Fernandes et al., 2016). Wang et al. quantify the impacts of increased shipping costs from BWM compliance on big economies, international trade, and global shipping traffic (Wang et al., 2020a). Cooper et al. considers possible water treatment methods that can be used in ballast water treatment, providing potential alternative methods for the control of invasive species currently being tested for use on ships (Cooper et al., 2007). There is also research that investigates various water treatment methods (Saleh, 2021, Gupta et al., 2012, Saleh, 2020).

There is no research that examines the impact of ballast water regulation on Small Island Developing States (SIDS) and Least Developed Countries (LDCs) to the best of our knowledge. However, SIDS and LDCs have unique social, economic, and environmental conditions, and are vulnerable to external shocks. SIDS and LDCs rely heavily on the world market, on one hand, to export oil, ores, phosphates, and other raw materials, which are important components of their national economies. On the other hand, SIDS are also disproportionately dependent on inter-island shipping for the delivery of necessary food, energy, and equipment imports (GloFouling Project, IMO; United Nations). Moreover, a relatively small number of ships are domestically owned by SIDS and LDCs, so these States largely rely on foreign-sourced shipping (UNCTADSTAT; ISWG-GHG-7-2-20). Shipping connectivity may be affected by cost changes resulting from BWM compliance and may disrupt trade under conditions where SIDS and LDCs do not operate or have ownership control of ocean transport fleets serving their economies. Therefore, this work evaluates the burden of ballast water regulation among trading nations, and in particular burden borne by SIDS and LDCs. We not only examine the economic impacts, but also study if the invasion risk of SIDS and LDCs are reduced to the same extent as bigger economies.

Many studies attempt to quantify the invasion risk caused by ballast water discharge. Some researchers establish global biological invasion models based on ship trajectories (Drake and Lodge, 2004, Kaluza et al., 2010, Keller et al., 2011), yet these models do not consider environmental and biogeographical factors (Seebens et al., 2013). Seebens et al. further consider survival and establishment pressures of alien species (Seebens et al., 2013). This model has been used to predict invasion risks (Seebens et al., 2016, Sardain et al., 2019); however, this model implicitly assumes that the invasion risk to a port only comes from the last port of call. Xu et al. further construct a higher-order network based on the first-order invasion risk model of Seebens et al. to capture the invasion risk including several previous ports of call (Xu et al., 2016, Saebi et al., 2020b). Accordingly, we use the model of Xu et al. and Saebi et al. to estimate the invasion risk to SIDS and LDCs before and after the BWM Convention.

A research gap exists regarding the analysis of potential impacts on SIDS and LDCs of the BWM Convention, and our work is the first to evaluate if the Convention generates inequality for different countries/regions, to the best of our knowledge. Also, there are no widely used metrics proposed or established that can be used to quantify inequity generated by shipping environmental regulations. The economic and environmental analysis results from this work provide the assessment of inequality of regulation for States and the IMO to work together to implement the BWM Convention in a fair and effective way. The metrics and methods of disproportionate impact assessment we employ can also be applied as an equity test for future research on other shipping environmental regulations including shipping decarbonization measures.

2. **Methods**

This research evaluates the environmental and economic impacts of the global IMO BWM Convention. To quantify changes in NIS risks, we calculate national-level NIS risks before and after ballast water treatment with a higher-order species flow model (SF-HON). For the economic impacts, we simulate the shipping cost change caused by the Convention in a computable general equilibrium (CGE) model. We then use the Lorenz Curve and Gini Coefficient to examine inequality effects of BWM impacts.

**2.1 Higher-order species flow model**

The risk assessment in this work is based on a Species Flow Higher-Order-Network (SF-HON) to consider higher-order species introduction pathways (Xu et al., 2016, Saebi et al., 2020a). In the context of ballast water discharge, ships may not discharge all ballast water at one single port and may discharge the remaining ballast water to the next calling ports. This will introduce the alien species from the original ports to many other ports. The SF-HON captures such ballast water discharge profiles and yields more accurate results compared to the First-Order-Network (Saebi et al., 2020b).

The SF-HON comprises three parts: the likelihood of a species being nonindigenous, the introduction probability, and the establishment probability (Seebens et al., 2013). Equation 1 defines the NIS risk between port $i$ and port $j$ of vessel $v$.

$$\text{P(NIS spread)}_{ij}^{v} = \text{P(nonindigenous)}_{ij} \times \text{P(intro)}_{ij}^{v} \times \text{P(establish)}_{ij} \quad \text{(Equation 1)}$$

The likelihood of being nonindigenous is 0 if both source and destination ports are in the same or neighboring ecoregions, and 1 otherwise.

$$P(\text{intro})_{ij}^v = \rho^v(1 - e^{-\lambda D_{ij}^v})e^{-\mu \Delta t_{ij}^v} \quad \text{(Equation 2)}$$

The introduction rate is measured with Equation 2, where $D_{ij}^v$ is the amount of ballast water discharged at a port; $\lambda = 3.22 \times 10^{-6}$, is the species introduction potential per volume of the discharge; $\mu = 0.02$, is the daily mortality rate of species in ballast water during the voyage, $\Delta t_{ij}^v$ is the duration of one shipping voyage; $\rho^v$ measures ballast water treatment efficacies and $1-\rho^v$ is the efficacy of BWM for the route. Ballast water treatment efficacy is estimated with the average species concentration in untreated ballast water and the concentration required in the IMO BWM Convention, which is 99.15%. The average species concentrations in untreated ballast water are from experimental results from the literature (Cohen and Dobbs, 2015, Reusser et al., 2013, Minton et al., 2005, Briski et al., 2015, Briski et al., 2014, Briski et al., 2013).

The establishment rate depends on the environmental similarity of the origin and destination ports in Equation 3.

$$P(\text{establish})_{ij} = \alpha e^{-\frac{1}{2}\left[\left(\frac{\Delta T_{ij}}{\delta T}\right)^2 + \left(\frac{\Delta S_{ij}}{\delta S}\right)^2\right]} \quad \text{(Equation 3)}$$

where $\Delta T_{ij}$ and $\Delta S_{ij}$ are the temperature and salinity difference of the origin and destination ports; $\delta T$ and $\delta S$ are the standard deviations in temperature and salinity, respectively, where $\delta T = 2\,°C$ and $\delta S = 10\,ppt$.

The NIS risks are aggregated over all routes $r$ of a ship $v$ from port $i$ to port $j$ to get the risk between every pair of ports with Equation 4.

$$P(\text{NIS spread})_{ij} = 1 - \Pi_{r,v}(1 - P(\text{NIS spread})_{ij}^v) \quad \text{(Equation 4)}$$

We feed these paths and their probabilities to the HON algorithm (Xu et al., 2016) to obtain the higher-order network of species flow, and build the physical adjacency matrix in which each edge weight (NIS spread risk) is calculated by averaging over all HON edges that correspond to that pair of ports. The edge weights are normalized by dividing all the edge weights by the maximum value in the network. Then we obtain the cumulative risk for each port $k$ by aggregating the NIS spread risk over all incoming ports $i$ with Equation 5 and calculate the average NIS risks of all ports in each region.

$$P(\text{NIS spread})_k = 1 - \Pi_i(1 - P(\text{NIS spread})_{ik}) \quad \text{(Equation 5)}$$

**2.2 Integrated CGE and shipping cost modeling framework**

We use an integrated CGE and shipping cost modeling approach to study the global economic impacts of ballast water management (Wang et al., 2020a). The modeling framework uses a shipping cost model to estimate the inputs used as exogenous shocks to the global shipping sector in the CGE model. The CGE model used is known as GTAP (Global Trade and Analysis Project), which is effective to study the impacts of different policies on global trade (Corong et al., 2017). We simulate the changes in international trade and other economic indicators by implementing the shipping cost changes as exogenous shocks in the economic model. The first step in this approach is to calculate the changes in shipping costs associated with the BWM Convention, and the second step is to simulate the economic impacts resulting from the changes in shipping costs.

**The CGE model.** The GTAP model has been used widely to examine the impacts of various policies on international trade and the environment (Beckman and Countryman, 2021, Bekkers et al., 2016, Countryman et al., 2016, Wang et al., 2020b, Countryman et al., 2018, Hertel and de Lima, 2020, Nong et al., 2019). CGE models include interactions between producers, consumers, investors, households, and governments, and are useful to characterize linkages between sectors and to investigate policies that have economy-wide impacts. Both instances occur when considering policies targeting transportation and shipping sectors including the IMO BWM Convention we investigate. We provide a brief description of the CGE model and specific changes made for this analysis are described fully. Full documentation of the GTAP model structure can be found in (Corong et al., 2017).

Economic agents in the model include households, government, industries, and investors. Households and governments are the final consumers in each country, purchasing goods and services produced domestically and internationally. Households maximize utility subject to their budget constraints. A household receives income by supplying labor, and the government receives tax revenue. Producers provide output for final consumption and for use as inputs to produce goods in other sectors. Each sector uses inputs for production that are sourced domestically and from international sources, labor, capital, and other resources. National economies are linked by bilateral trade. The GTAP model includes transportation services used by producers to move goods between sectors and countries.

**Shipping cost model.** In the context of this work, the exogenous economic shock to the CGE model is the change in maritime transportation costs due to ballast water treatment required by regulation. The change in the shipping costs are determined by using two models: (1) the regulatory compliance cost model and (2) the baseline shipping cost model (Wang et al., 2020b). The regulatory compliance cost model used in this work is established by Wang and Corbett (Wang and Corbett, 2020). The aforementioned authors modeled the compliance costs including capital, instillation, and treatment of using BWMS to achieve the standards of both the IMO Convention and California standards. The compliance cost for each voyage is obtained with Equation 6.

$$(C + O)/N_v + T * V_v \qquad \text{(Equation 6)}$$

Where $C$ and $O$ are the annual capital and operating costs of BWMS ($); $N_v$ is the number of treatments of vessel $v$ in one year; the capital cost is assumed to be shared by all treatments of vessel $v$; $T$ is the treatment cost for each tonnage of ballast water ($/ton); $V_v$ is the ballast water treatment volume of vessel $v$ for that voyage (ton).

Each country/region pair has more than one shipping voyage in one year and we aggregate the costs of all voyages between each country/region pair to get annual costs.

The baseline shipping costs are calculated by including voyage duration and daily shipping costs. Daily shipping costs are derived from the Guide to Deep-Draft Vessel Operating with a 2002 base year (US Army Corps of Engineers, 2002), following the literature (Corbett et al., 2009, Dumas and Whitehead, 2008, The EPA, 2012). The daily shipping costs were estimated separately for containerships, bulkers, and tankers. Voyage durations are calculated by using shipping voyage information from Lloyd's List Intelligence (See data description in Section 2.4). The method and data are verified by Wang et al. to reflect current costs (Wang et al., 2020a).

**Studied countries/regions.** The GTAP 10 Data Base we use includes 141 regions[1] comprised of individual countries and regional groupings of countries if there is no country-level social accounting matrix available (Aguiar et al., 2019). There are 58 countries identified as SIDS and 46 countries are classified as LDCs as of December 21, 2020[2].

---

[1] https://www.gtap.agecon.purdue.edu/databases/regions.aspx?version=10.211

We study the countries that face the highest compliance costs for which we have data. We include 23 States and aggregated regions in our analysis: 10 SIDS and LDCs[3], 2 large traders among SIDS and LDCs (Brazil and Indonesia), the 2 largest traders in the world (USA and China), 3 aggregated regions of SIDS and LDCs (Oceania SIDS and LDCs, Rest of the Caribbean SIDS and LDCs, and Rest of the African SIDS and LDCs), and 6 other geographic regions[4]. The SIDS or LDCs not included separately are a result of data availability and are aggregated together according to geographic regions. The region aggregational was determined by the following steps.

The first step is the selection of separate States. The latest GTAP Database 10A includes 7 of 58 SIDS and 17 of 46 LDCs separately. We use the Liner Shipping Connectivity Index (LSCI) to represent the reliance of SIDs and LDCs on trade and/or shipping (Fugazza and Hoffmann, 2017). We found that 7 SIDS and 9 LDCs have LSCI values and we selected the top 5 SIDS and 5 LDCs with the highest LSCI as representative in this analysis (See Footnote 3). We also include Indonesia and Brazil, large trade partners of these 10 countries with the highest shipping traffic for exports to SIDS and LDCs, and the U.S. and China, the two largest traders in the world.

We include three more aggregate regions composed of SIDS and LDCs: Oceania SIDS and LDCs, Rest of the Caribbean SIDS and LDCs, and Rest of the African SIDS and LDCs. These three regions are based on the aggregate regions in the GTAP database and are primarily SIDS and LDCs. Table A1 shows how we determine the regions in our analysis from the GTAP database with the number of SIDS and LDCs indicated. Other regions (including those incorporating several SIDS and LDCs) are aggregated into regions described in Footnote 4. Table A2 provides the abbreviations for these States and regions.

**Sector aggregation**. The GTAP Database 10A includes 65 separate sectors that we aggregate into 25 groups based on the dependence of SIDS and LDCs on these sectors (See Table S1 in the Supplementary Data). We leave detailed food commodities separate, given the importance of food security in SIDS and LDCs. Also, energy sectors are kept separate, such as coal, crude oil, petroleum products, and natural gas because energy sectors are either key imports or exports in many SIDS and LDCs. Other important sectors include textiles, chemicals, metal and minerals, and machinery and equipment.

## 2.3 Inequality measurements: Lorenz curve and Gini coefficient

The Lorenz curve was initially used to study inequality in income distribution in human populations (Lorenz, 1905) and the traditional Gini coefficient focused on measuring income distribution (Ruitenbeek, 1996, Blackwood and Lynch, 1994). The Lorenz curve and Gini coefficient have also been used to examine expenditure inequality (Goodman and Oldfield,

---

[2] https://www.un.org/ohrlls/content/list-sids
https://www.mfa.gov.sg/SINGAPORES-FOREIGN-POLICY/International-Issues/Small-States (Singapore)
https://www.un.org/ohrlls/content/profiles-ldcs
[3] Singapore, Dominican Republic, Jamaica, Mauritius, Bahrain, Togo, Benin, Senegal, United Republic of Tanzania, and Bangladesh
[4] Europe, Rest of Asia, Rest of Africa, Rest of North America, Rest of South America, and Oceania and Rest of World

2004), inequality in energy demand (Papathanasopoulou and Jackson, 2009), and inequality in energy consumption (Jacobson et al., 2005). Broader applications of inequality of environmental aspects include climate change impacts (Tol et al., 2004, Hedenus and Azar, 2005), resource use, emissions resulting from resource use and solid waste arising from material resource use (Druckman and Jackson, 2008). The Lorenz curve can also be applied to issues of inequality measurement of carbon emission abatement policies (Groot, 2010) and the Paris Agreement (Zimm and Nakicenovic, 2020). When measuring the international equality of reducing emissions from deforestation, Cattaneo et al. propose two ways to describe equality: relative to the endowment of carbon and relative to total opportunity cost (Cattaneo et al., 2010).

This work applies the Lorenz curve and Gini coefficient to the issue of inequality of impacts from the global BWM Convention. We use two ways to describe equality: first, in terms of invasion risk reduction, and second, economic impacts measured as risk reduction relative to income levels and changes in economic indicators relative to income levels, where income is measured as per capita GDP. In our work, if the Lorenz curve matches the 45-degree line, it means the full set of nations including SIDS and LDCs are equally impacted by the IMO BWM. The Gini coefficient values lie between 0 to 1, where 0 represents absolute equality and 1 represents absolute inequality (Ruitenbeek, 1996, Druckman and Jackson, 2008).

**2.4 Data**

NIS risks are estimated by employing ship movement data, environmental data, ballast water discharge profiles, and ballast water treatment efficacy data of BWMS. Ship movement data are from Lloyd's List Intelligence (2018-2019), with ship voyage information including ports, sail date, arrival date, and vessel specifications (deadweight tonnage, vessel type, etc.). Only ship movements involved in international shipping are used since the BWM Convention regulates international ships. Ballast water discharge volume and frequency are estimated with data from National Ballast Information Clearinghouse (Saebi et al., 2020a). Environmental data (temperature and salinity) are from the Global Ports Database (Keller et al., 2011), the World Ocean Atlas (Zweng et al., 2013, Locarnini et al., 2013), and Marine Ecoregion of the World (Spalding et al., 2007).

Besides the shipping movement data and ballast water discharge profiles, the coupled shipping cost and CGE model use costs of BWMS, shipping costs, baseline economic data of regions, and behavioral parameters associated with production, consumption, and trade. The capital and operating costs of BWMS are from King et al. (King et al., 2009). Costs vary for different BWMS given their different treatment capacity and treatment methods. For example, the capital cost of a BWMS can vary between $0.2-1 million. In this work, we use the average BWMS costs to calculate the Convention compliance cost. The capital and installation cost used is $0.9 million, which is about $49,000 per year. The annual operating cost is $13,500. The cost to treat one ton of ballast water is $0.135. Shipping costs are from the U.S. Army Corps of Engineers (US Army Corps of Engineers, 2002), including daily capital, operating, and fuel costs. The baseline economic data and behavioral parameters are from GTAP Database version 10A (Aguiar et al., 2019). Specifically, we employ our estimated changes in shipping costs resulting from the implementation of IMO BWM Convention regulation as exogenous shocks in the economic model.

**3  Results and discussion**

We present the impacts of the IMO BWM Convention from four aspects: invasion risk, shipping cost, economic indicators, and bilateral trade.

**3.1 Invasion risk reduction**

This section examines (1) the efficacy of the IMO BWM Convention to reduce invasion risk; (2) whether inequality in NIS risk reduction exists among all regions; and (3) if the IMO BWM Convention generates disproportionate environmental impacts on SIDS and LDCs compared to more developed countries and regions.

Table 3-1 shows the NIS risks of each region with and without the IMO BWM and risk reduction ratio by the IMO BWM. The NIS risks described are the average risk of ports within each region. Risk reduction ratios of all regions are above 99%, except Singapore. In all cases, except Singapore, the risk reduction due to modeled compliance with the IMO BWM Convention is about two orders of magnitude (ranging from a risk reduction ratio of 118 to 273, except for Singapore with a risk reduction ratio of 22). Such results illustrate the effectiveness of the Convention.

Table 3-1 NIS risks and risk reduction of regions (ranked by risk reduction ratio)

|    | Region | NIS risk with no policy | NIS risk under IMO BWM | Risk reduction (ratio) | Risk reduction (order of magnitude) |
|----|--------|-------------------------|------------------------|------------------------|--------------------------------------|
| 1  | Singapore | 0.29426 | 0.01345 | 95.43% | 22 |
| 2  | Mauritius | 0.00378 | 0.00003 | 99.15% | 126 |
| 3  | Benin | 0.02658 | 0.00017 | 99.36% | 156 |
| 4  | Senegal | 0.07133 | 0.00046 | 99.36% | 155 |
| 5  | Tanzania | 0.01266 | 0.00008 | 99.36% | 158 |
| 6  | Bangladesh | 0.10934 | 0.00062 | 99.43% | 176 |
| 7  | Jamaica | 0.02602 | 0.00015 | 99.44% | 173 |
| 8  | USA | 0.01307 | 0.00007 | 99.47% | 187 |
| 9  | Brazil | 0.03398 | 0.00018 | 99.47% | 189 |
| 10 | China | 0.05510 | 0.00028 | 99.49% | 197 |
| 11 | Togo | 0.02052 | 0.00012 | 99.53% | 171 |
| 12 | Rest North America | 0.02237 | 0.00011 | 99.53% | 203 |
| 13 | Europe | 0.01261 | 0.00006 | 99.55% | 210 |
| 14 | Caribbean SIDS | 0.03045 | 0.00014 | 99.55% | 218 |
| 15 | Oceania and Rest of World | 0.04885 | 0.00022 | 99.56% | 222 |
| 16 | India | 0.13925 | 0.00059 | 99.57% | 236 |
| 17 | Oceania SIDS | 0.01271 | 0.00005 | 99.58% | 254 |
| 18 | Rest Asia | 0.05397 | 0.00023 | 99.58% | 235 |
| 19 | Rest Africa | 0.05401 | 0.00022 | 99.59% | 246 |
| 20 | Bahrain | 0.02950 | 0.00012 | 99.59% | 246 |
| 21 | Dominican Republic | 0.02725 | 0.00010 | 99.63% | 273 |
| 22 | Rest South America | 0.02448 | 0.00012 | 99.63% | 204 |
| 23 | African SIDS and LDCs | 0.03141 | 0.00012 | 99.42% | 262 |

Note: NIS risks with and without BWM are determined by the higher-order species flow model. NIS risk under the IMO BWM Convention is simulated in the HON model with ballast water treatment efficacy of 99.15%, assuming the numeric organism concentration required in the Convention can be fully achieved. The NIS risk ratio is calculated by: (NIS risk with no policy - NIS risk under IMO BWM)/ NIS risk with no policy.

Environmental equality is measured by NIS risk reduction relative to income level (GDP per capita). Figure 3-1 is the Lorenz Curve with the Gini coefficient equal to 0.66. This shows that inequality exists in the distributional effects of NIS risk changes relative to per capita income. However, from Table 3-1 we can see some SIDS and LDCs are at higher or lower risk than the USA, Europe, China, Rest Africa, etc. This means the patterns of risk and risk-reduction among countries are not observably different among SIDS/LDCs and other economies.

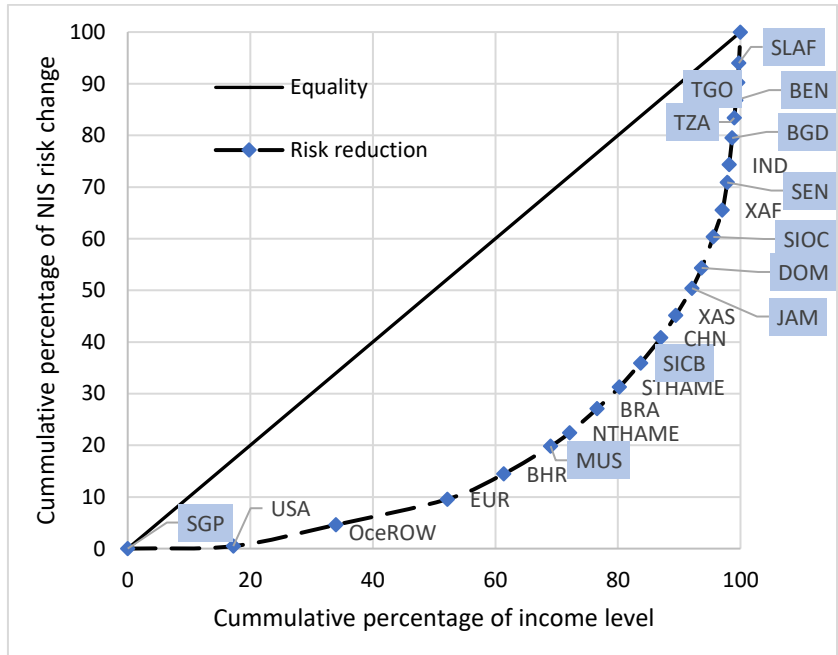

Figure 3-1 The Lorenz Curve of NIS risk reduction and income level of regions. SIDS and LDCs are shown in the frames. The Gini coefficient is 0.66.
Source: Authors' simulation

### 3.2 Changes in shipping costs

Changes in shipping costs are estimated as the compliance costs of the IMO BWM Convention. Table 3-1 shows the percentage changes in shipping costs between regions. Results show that shipping costs for voyages between most regions do not change substantially (less than 5%). However, shipping costs between some trading partners do change greatly, and the bigger changes (larger than 5%) are highlighted in Table 3-1. One finding is that all these relatively large shipping cost changes occur for trade to or from SIDS and LDC countries, while there are no big cost changes for shipments between developed countries/regions. There are some trade partners with cost changes larger than 10% and the most affected regions are Togo, Bangladesh, Jamaica, Caribbean SIDS, and Oceania SIDS. For example, the shipping cost of voyages from Bangladesh to Oceania SIDS increases by 27%, and from China to Togo increases by 21%. These relatively large shipping cost changes are due to high BWM compliance costs from the use of BWMS relative to their baseline shipping costs. The absolute changes in shipping costs (compliance costs) can be found in Table S2 in the Supplementary Data.

Table 3-1 Changes in shipping costs from IMO BWM Convention Compliance

| Region | BEN | BGD | BHR | BRA | CHN | DOM | EUR | IND | JAM | MUS | North Ame | Oce ROW | SEN | SGP | SICB | SIOC | SLAF | South Ame | TGO | TZA | USA | XAF | XAS |
|---|---|---|---|---|---|---|---|---|---|---|---|---|---|---|---|---|---|---|---|---|---|---|---|
| BEN | | | | 1% | | | 0% | | | | | | 2% | | | 2% | 2% | 1% | 8% | | | 2% | 0% |
| BGD | | | | 1% | 2% | | 4% | 6% | | 3% | 3% | 2% | 0% | 2% | 18% | 27% | 2% | 1% | | | 3% | 1% | 2% |
| BHR | | 5% | | 1% | 1% | | 4% | 2% | | | 1% | 6% | | 1% | | | 1% | 1% | | | | 1% | 3% |
| BRA | | 1% | 1% | | 0% | 1% | 1% | 1% | 1% | 1% | 1% | 1% | 1% | 1% | 1% | 3% | 1% | 2% | 1% | | 1% | 1% | 1% |
| CHN | | 2% | 2% | 1% | | | 2% | 2% | | 1% | 1% | 1% | | 1% | 7% | 2% | 1% | 1% | 21% | 2% | 1% | 0% | 2% |
| DOM | | | | 0% | | | 0% | | 2% | | 2% | | | 0% | 2% | | | 3% | | | 2% | 0% | 0% |
| EUR | 1% | 2% | 3% | 1% | 1% | 0% | 3% | 2% | 1% | 3% | 1% | 2% | 1% | 1% | 1% | 2% | 1% | 1% | 1% | 1% | 1% | 1% | 2% |
| IND | 2% | 4% | 2% | 1% | 1% | 1% | 2% | | 28% | 2% | 1% | 2% | 1% | 2% | 1% | 1% | 1% | 1% | 1% | 0% | 1% | 1% | 2% |
| JAM | | 0% | | 1% | 0% | 2% | 1% | 28% | | | 2% | | | 1% | 5% | | | 2% | | | 2% | 1% | 1% |
| MUS | | 2% | | 1% | 1% | | 2% | 2% | | | 0% | 0% | | 1% | | 1% | 1% | 1% | 0% | 1% | 0% | 1% | 1% |
| NTHAME | | 2% | 0% | 1% | 1% | 2% | 1% | 0% | 2% | 1% | | 2% | 1% | 1% | 1% | 2% | 1% | 1% | 2% | | 2% | 1% | 0% |
| OceROW | | 1% | 4% | 1% | 1% | | 1% | 1% | | 1% | 1% | 2% | | 1% | 1% | 2% | 1% | 1% | 0% | 1% | 1% | 1% | 2% |
| SEN | 2% | | | 2% | | | 1% | 1% | | | 1% | | | 1% | | 2% | 5% | 1% | 1% | | 1% | 1% | 2% |
| SGP | 0% | 2% | 1% | 1% | 1% | | 1% | 2% | | 1% | 0% | 1% | 0% | | 0% | 1% | 0% | 0% | 1% | 1% | 0% | 0% | 2% |
| SICB | | 0% | | 2% | 0% | 3% | 1% | 0% | 4% | 1% | 2% | 1% | 2% | 0% | 4% | | 0% | 3% | 1% | | 2% | 0% | 0% |
| SIOC | 1% | 5% | | 2% | 1% | | 1% | 1% | 1% | 1% | 1% | 2% | 4% | 1% | | 3% | 2% | 2% | 1% | | 1% | 1% | 1% |
| SLAF | 2% | 2% | 2% | 1% | 0% | | 1% | 1% | 2% | 1% | 1% | 2% | 4% | 0% | | 3% | 2% | 1% | 2% | 1% | 1% | 2% | 1% |
| STHAME | | 1% | 1% | 2% | 0% | 2% | 1% | 0% | 2% | 1% | 2% | 1% | 1% | 0% | 2% | 1% | 1% | 2% | 1% | 1% | 1% | 1% | 1% |
| TGO | 7% | | | 1% | 1% | | 1% | 1% | 1% | 1% | 1% | 1% | 1% | 0% | 1% | 2% | 2% | 1% | | | 1% | 3% | 0% |
| TZA | | | | 1% | 0% | | 1% | 1% | | 1% | | 1% | | 1% | | | 2% | 1% | | | | 3% | 1% |
| USA | 1% | 1% | | 1% | 0% | 1% | 1% | 0% | 1% | 0% | 2% | 2% | 1% | 0% | 2% | 1% | 1% | 1% | 1% | | | 0% | 0% |
| XAF | 3% | 1% | 1% | 1% | 0% | 1% | 1% | 1% | 1% | 1% | 1% | 1% | 1% | 1% | 1% | 2% | 2% | 1% | 4% | 4% | 0% | | 1% |
| XAS | 1% | 2% | 3% | 1% | 2% | 1% | 2% | 2% | 0% | 1% | 0% | 2% | 1% | 2% | 1% | 2% | 2% | 1% | 0% | 1% | 1% | 1% | 2% |

Note: Percentages are rounded to integrals for readability. Blanks indicate no shipping traffic is observed for the route. Full names of countries and regions are in Table A2.
Source: Authors' calculations

## 3.3 Economic effects of the IMO BWM Convention

All macroeconomic impacts due to the IMO BWM regulation are negative, but the overall effects of IMO BWM compliance on national economies are minor. None of the effects on national GDP, real exports, or real imports are substantial. Most changes are no larger than 0.1% for all countries/regions in the analysis (Table 3-2). Results can be explained by the small cost shocks, in general, relative to the sizes of national economies. One exception is Togo, given that Togo is the most affected by the regulation with corresponding decreases in GDP, exports, and imports of 0.8%, 0.9%, and 0.7%, respectively with the implementation of BWM standards. The economic effects of IMO BWM compliance in this study are consistent with findings from research examining the economic effects of the IMO regulation on big economies (Wang et al., 2020b).

The effects on economic welfare have similar patterns. The global implementation of IMO BWM regulation negatively affects economic welfare[5]. Welfare losses are less than 0.1% for all countries/regions, except Togo, as indicated in Table S3 in the Supplementary Data. Togo is most negatively affected by higher transport costs with economic welfare decreases by $45 million (1.13%) resulting from IMO BWM regulation. Similarly, the overall changes of sectoral outputs, imports, and exports are also relatively small as shown in Tables S4-S6 in the Supplementary Data. Sectoral changes of most regions are less than 0.5%. Again, Togo is the most affected country with agriculture imports decreased by 1.3% and coal export declined by 1.4%.

Table 3-2 Macroeconomic changes of regions

| Region | GDP | | Export | | Import | |
| --- | --- | --- | --- | --- | --- | --- |
| | Percentage | Real change ($ million) | Percentage | Real change ($ million) | Percentage | Real change ($ million) |
| Togo | -0.8% | -3362 | -0.9% | -2015 | -0.7% | -3632 |
| Bangladesh | -0.1% | -8644 | -0.2% | -7933 | -0.3% | -12752 |
| Benin | -0.0% | -383 | -0.1% | -97 | -0.1% | -471 |
| Jamaica | -0.0% | -557 | -0.1% | -308 | -0.1% | -562 |
| Oceania SIDS and LDCs | -0.0% | -2038 | -0.1% | -932 | -0.1% | -1964 |
| Tanzania | -0.0% | -1921 | -0.1% | -454 | -0.1% | -1464 |
| Senegal | -0.0% | -470 | -0.1% | -416 | -0.1% | -849 |
| Bahrain | -0.0% | -677 | -0.1% | -1238 | -0.1% | -2049 |
| Dominican Republic | -0.0% | -1279 | -0.1% | -719 | -0.1% | -1707 |
| Europe | -0.0% | -442720 | -0.1% | -485465 | -0.0% | -407400 |
| Mauritius | -0.0% | -252 | -0.1% | -307 | -0.1% | -480 |
| Oceania and ROW | -0.0% | -33103 | -0.1% | -20400 | -0.1% | -31954 |
| Caribbean SIDS and LDCs | -0.0% | -5098 | -0.1% | -6277 | -0.1% | -9268 |
| African SIDS and LDCs | -0.0% | -3776 | -0.0% | -2388 | -0.1% | -3965 |
| Rest Asia | -0.0% | -263800 | -0.1% | -360926 | -0.1% | -434741 |
| Brazil | -0.0% | -24170 | -0.1% | -15671 | -0.1% | -19010 |

---

[5] We measure economic welfare in terms of equivalent variation, a money-metric measure of economic well-being associated with changes in prices.

| | | | | | | |
|---|---|---|---|---|---|---|
| China | -0.0% | -106423 | -0.1% | -211429 | -0.1% | -203541 |
| India | -0.0% | -20424 | -0.1% | -37601 | -0.1% | -36339 |
| Rest North America | -0.0% | -30897 | -0.0% | -18493 | -0.0% | -30511 |
| Singapore | -0.0% | -3063 | -0.1% | -30128 | -0.0% | -6579 |
| Rest South America | -0.0% | -23490 | -0.1% | -22723 | -0.1% | -34459 |
| Rest Africa | -0.0% | -21804 | -0.0% | -16938 | -0.0% | -25726 |
| USA | 0.0% | 0 | -0.0% | -39216 | -0.0% | -52159 |

Note: Changes in absolute trade value are round to integrals. Percentage changes are rounded to one decimal
Source: Authors' simulation

We further examine economic equality using GDP as an illustration. Figure 3-2 shows Lorenz curves of cumulative GDP changes relative to the cumulative income level of regions, with a Gini coefficient of 0.58. Results indicate that regional economies are affected to different extents relative to regions' income levels. For example, Europe may experience the largest extent of GDP decrease under the IMO regulation, with a 10% income level. However, no evidence shows inequality based on development status.

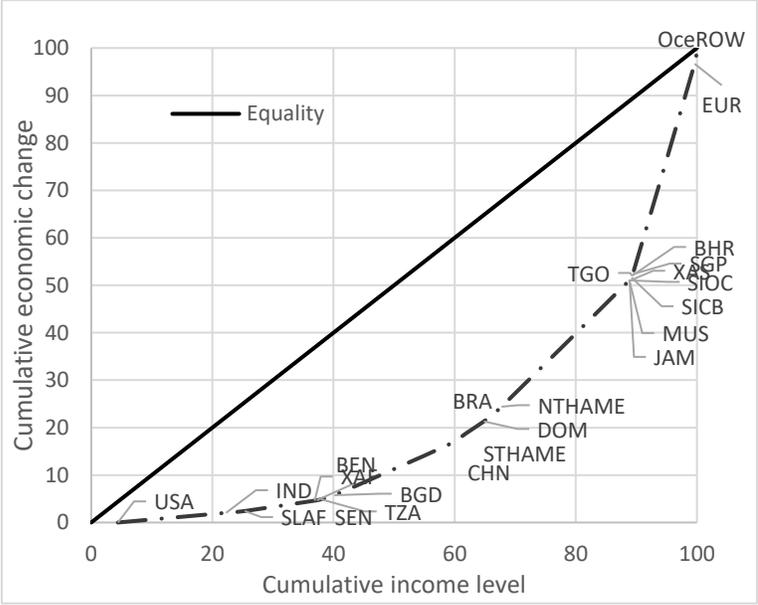

Figure 3-2 Inequality of GDP changes related to the income level. The Gini coefficient is 0.58.
Source: Authors' simulation

### 3.4 Bilateral trade changes

Disruptions in bilateral trade caused by the BWM Convention are not substantial for most product categories, except for trade in some sectors for several trade partners. The larger bilateral trade changes may happen for SIDS and LDCs, their corresponding largest trader partners, and other large economies for select sectors. The most affected countries include Togo, Bangladesh, Dominican Republic, India, and China. Table 3-3 provides changes in the bilateral commodity trade. Considering the small original export values of some sectors, we only include the bilateral trade with larger changes determined by percentage changes greater than 0.5% and absolute value changes greater than $1 million for at least one commodity for each trade partner as described in the table.

Togo, Bangladesh, and Dominican Republic experience export decrease in textiles and metal and chemicals. For example, Togo's exports of metal and chemicals to Rest Africa, India, and Rest Asia are simulated to decrease by 1.8% ($2 million), 1.0% ($1.6 million), and 0.8% ($3.6 million), respectively. Exports of textiles, metal and chemicals, and other agriculture from Bangladesh to India decline by 1.1% ($2.4 million), 2% ($1.7 million), and 2.3% ($1.9 million). Substitution effects are expected for Indian trade. For example, India's exports of textiles and metal and chemicals to Bangladesh decrease, while India's exports of these commodities to Togo increase. There are some noteworthy changes to exports from China. For example, China's exports of textiles, metal and chemicals, machinery equipment, and other agriculture to Togo and Caribbean SIDS experience large decreases from 0.6% to 6.9%, respectively. This is consistent with the substantial increases in shipping costs from China to Togo and Caribbean SIDS, which are 21% and 7%, respectively.

Table 3-3 Changes in exports by commodity between countries

| Commodities | Togo to | | | | | | Bangladesh to | | Dominican Republic to | |
|---|---|---|---|---|---|---|---|---|---|---|
| | Rest Africa | | India | | Rest Asia | | India | | Rest South America | |
| | (a) | (b) | (a) | (b) | (a) | (b) | (a) | (b) | (a) | (b) |
| Textiles | -3.6% | -2 | -3.1% | 0 | -3.0% | 0 | -1.1% | -2 | -1.2% | 0 |
| Metal& chemicals | -1.8% | -2 | -1.0% | -2 | -0.8% | -4 | -2.0% | -2 | -0.5% | -1 |
| Other agriculture | -0.9% | 0 | -0.2% | 0 | -0.2% | 0 | -2.3% | -2 | -0.4% | 0 |
| Machin equipment | -1.4% | -1 | -1.3% | 0 | -1.3% | 0 | -0.3% | 0 | -0.2% | 0 |
| Gas | 0.4% | 0 | 4.4% | 0 | 4.1% | 0 | 5.2% | 0 | -0.5% | 0 |

| Commodities | India to | | | | China to | | | |
|---|---|---|---|---|---|---|---|---|
| | Bangladesh | | Togo | | Togo | | Caribbean SIDS | |
| | (a) | (b) | (a) | (b) | (a) | (b) | (a) | (b) |
| Textiles | -1.1% | -16 | 4.0% | 5 | -2.5% | -26 | -1.0% | -10 |
| Metal& chemicals | -1.1% | -14 | 2.9% | 2 | -6.9% | -22 | -1.3% | -16 |
| Other agriculture | -0.9% | -10 | -0.1% | 0 | -5.5% | -4 | -0.9% | -1 |
| Machin equipment | -0.7% | -6 | 2.2% | 1 | -0.9% | -4 | -0.6% | -20 |
| Gas | -6.7% | 0 | 3.6% | 0 | 3.4% | 0 | 2.0% | 0 |

Note: (a) indicates percentage change, (b) indicates value change ($ million). Changes in absolute trade value are round to integrals. Percentage changes are rounded to one decimal. Results include trade partners with both percentage changes greater than 0.5% and absolute value changes greater than $1 million for at least one commodity of each trade partner.
Source: Authors' simulations

## 4 Conclusion

This work examines the potential disproportionate impacts of the technological costs on SIDS and LDCs, which are vulnerable to both socio-economic and species invasion shocks. We include both environmental and economic metrics: invasion risk reductions, economic impacts, and bilateral trade changes. In general, the IMO BWM Convention decreases biological invasion risk. The negative economic impacts of the Convention are minor across countries. This shows the feasibility of the IMO BWM Convention while considering the potential environmental and economic impacts on SIDS and LDCs. However, special attention is warranted for exports of textiles and metal and chemicals from Togo, Bangladesh, and Dominican Republic, which are more negatively affected than exports of products in other sectors across countries.

The Lorenz Curve and Gini coefficient reveal inequality exists for environmental and economic effects among nations; illustrated by the Gini coefficients equal to 0.66 and 0.58, respectively. However, we do not find that SIDS and LDCs are systematically different than large or developed economies given this risk assessment and economic analysis. The invasion risk reduction ratios of all regions are above 99%, with risk reduction ratios ranging from 118 to 273, with the exception of Singapore, which has a risk reduction ratio of 22. This demonstrates the strong protection efficacy of the BWM Convention. The Convention compliance cost changes for most regions are less than 5%, but some trade partners experience cost changes larger than 10%. For example, the shipping cost of voyages from Bangladesh to Oceania SIDS increases by 27%, and from China to Togo increases by 21%. In terms of economic impacts, none of the effects on national GDP, real exports, real imports, or economic welfare are large across countries/regions. The overall changes of sectoral outputs, imports, and exports are smaller than 0.1% for all countries/regions, except Togo, which is most affected (by about 1%). In terms of bilateral trade, disruptions are not substantial either, except for trade between several trade partners for some sectors. The larger bilateral trade changes are expected for SIDS, LDCs and bigger economies. Countries with the largest changes to bilateral trade patterns include Togo, Bangladesh, Dominican Republic, India, and China. For example, Togo's exports of metal and chemicals to Rest Africa, India, and Rest Asia are simulated to decrease by 1.8% ($2 million), 1.0% ($1.6 million), and 0.8% ($3.6 million), respectively. Exports of textiles, metal and chemicals, and other agriculture from Bangladesh to India decline by 1.1% ($2.4 million), 2% ($1.7 million), and 2.3% ($1.9 million), respectively. China's exports of textiles, metal and chemicals, machinery equipment, and other agriculture to Togo and Caribbean SIDS experience large decreases from 0.6% to 6.9%, respectively.

Results do not show that SIDS and LDCs are disproportionately impacted by the BWM Convention. We also see the strong protection from invasion risks due to the BWM Convention and minor economic and trade impacts resulting from the technological compliance strategies. This provides evidence to support the IMO BWM Convention from an environmental and economic perspective. However, what should be noticed is that the minor impacts on SIDS do not mean the effects are not important. Lower-income countries and households are especially vulnerable to any shock, even if relatively small, compared to high-income countries and households (Hertel et al., 2010). Further studies on the impacts of increased regulation on developing countries is warranted, and capacity building and international cooperation are still needed.

We admit the compliance costs to meet the standards of the IMO BWM Convention may vary due to the use of different BWMS. Costs of BWMS vary according to different treatment methods and treatment capacities. Also, due to potential technology advancements, the cost of BWMS may decrease over time. Our analysis using average costs may over or underestimate the changes in shipping costs, which may over or underestimate the economic and trade impacts. Our results show that the IMO BWM Convention does not generate large economic impacts (less than 0.5% change in economic impacts, in general) on SIDS/LDCs. Thus, if the costs of BWMS decrease in the future, the impacts on SIDS/LDCs remain small and our conclusions stand.

The methods we employ to investigate the environmental and economic equality effects of BWM policy can be applied to assess other regulations such as The International Convention for the Prevention of Pollution from Ships (MARPOL) and Biofouling management. Impact assessments can be completed better with more data availability. We use estimated shipping costs, instead of real freight rates to simulate the shocks for the CGE model since freight rates are difficult to obtain. Limited by the economic and trade data availability for many SIDS and LDCs, this work considers twelve SIDS and LDCs and includes others in aggregated regions (such as other SIDS and LDCs in Africa). More data is needed to provide increased country-specific analysis to evaluate the impacts on SIDS and LDCs around the world.

**CRediT authorship contribution statement**

**Zhaojun Wang:** Conceptualization, Methodology, Software, Validation, Formal analysis, Investigation, Data curation, Writing - original draft. **Amanda M. Countryman:** Conceptualization, Methodology, Resources, Writing - original draft, Supervision, Project administration, Funding acquisition. **James J. Corbett:** Conceptualization, Methodology, Resources, Writing - original draft, Supervision, Project administration, Funding acquisition. **Mandana Saebi**: Methodology, Software, Validation, Formal analysis, Investigation, Data curation

**Acknowledgments**

This work was supported by, "Changes in Ship-born Invasions in Coupled Natural-Human Ecosystems: Infrastructure, Global Trade, Climate, and Policy," P.I.: D. Lodge, CoPIs: N. Chawla, A.M. Countryman, E. Grey, National Science Foundation: Coastal SEES Collaborative Research, NSF Award Number: 17483892, Changes in Ship-born Invasions in Coupled Natural-Human Ecosystems: Infrastructure, Global Trade, Climate, and Policy," P.I.: J. Corbett, National Science Foundation: Coastal SEES Collaborative Research, NSF Award Number: 1426973, and by the USDA National Institute of Food and Agriculture, Hatch project COL00232A, accession 1016945.

**Appendix**

### (1) Region aggregation

Table A1. Aggregate regions containing GTAP with SIDS and LDCs

| Aggregate regions in the GTAP database | Number of States in GTAP | Number of SIDS | Number of LDCs | Regions used in this work |
|---|---|---|---|---|
| Rest of Oceania | 23 | 19 | 3 | Oceania SIDS and LDCs |
| Rest of Caribbean | 18 | 17 | 1 | Rest Caribbean SIDS and LDCs |
| Rest of Western Africa | 9 | 2 | 7 | Rest African SIDS and LDCs |
| Rest of Eastern Africa | 8 | 2 | 6 | Rest African SIDS and LDCs |
| Rest of South-Central Africa | 2 | 0 | 2 | Rest African SIDS and LDCs |
| Rest of Central Africa | 6 | 1 | 3 | Rest of Africa |
| Rest of Southeast Asia | 2 | 1 | 2 | Rest of Asia |
| Rest of South Asia | 3 | 1 | 2 | Rest of Asia |
| Rest of Western Asia | 20 | 0 | 1 | Rest of Asia |
| South America | 5 | 2 | 0 | Rest of America |
| North America | 3 | 1 | 0 | Rest of America |
| Central America | 1 | 1 | 0 | Rest of America |

Note: some States belong to both SIDS and LDCs
Source: Authors' summary

Table A2. Studied regions and abbreviations

| Abbreviation | Region |
|---|---|
| BEN | Benin |
| BGD | Bangladesh |
| BHR | Bahrain |
| BRA | Brazil |
| CHN | China |
| DOM | Dominican Republic |
| EUR | Europe |
| IND | India |
| JAM | Jamaica |
| MUS | Mauritius |
| NTHAME | North America |
| OceROW | Oceania and Rest of World |
| SEN | Senegal |
| SGP | Singapore |
| SICB | Caribbean SIDS |
| SIOC | Oceania SIDS |
| SLAF | African SIDS and LDCs |
| STHAME | South America |
| TGO | Togo |
| TZA | Tanzania |
| USA | USA |
| XAF | Rest Africa |
| XAS | Rest Asia |

# References

AGUIAR, A., CHEPELIEV, M., CORONG, E. L., MCDOUGALL, R. & VAN DER MENSBRUGGHE, D. 2019. The GTAP Data Base: Version 10. *2019,* 4**,** 27.

BECKMAN, J. & COUNTRYMAN, A. M. 2021. The Importance of Agriculture in the Economy: Impacts from COVID-19. *American Journal of Agricultural Economics*.

BEKKERS, E., FRANCOIS, J. F. & ROJAS-ROMAGOSA, H. 2016. Melting ice caps and the economic impact of opening the northern sea route. *The Economic Journal*.

BLACKWOOD, D. L. & LYNCH, R. G. 1994. The measurement of inequality and poverty: A policy maker's guide to the literature. *World Development,* 22**,** 567-578.

BRISKI, E., ALLINGER, L. E., BALCER, M., CANGELOSI, A., FANBERG, L., MARKEE, T. P., MAYS, N., POLKINGHORNE, C. N., PRIHODA, K. R. & REAVIE, E. D. 2013. Multidimensional approach to invasive species prevention. *Environmental Science & Technology,* 47**,** 1216-1221.

BRISKI, E., GOLLASCH, S., DAVID, M., LINLEY, R. D., CASAS-MONROY, O., RAJAKARUNA, H. & BAILEY, S. A. 2015. Combining ballast water exchange and treatment to maximize prevention of species introductions to freshwater ecosystems. *Environmental science & technology*.

BRISKI, E., LINLEY, R. D., ADAMS, J. & BAILEY, S. A. 2014. Evaluating efficacy of a ballast water filtration system for reducing spread of aquatic species in freshwater ecosystems. *Management of Biological Invasions,* 5**,** 245-253.

CARLTON, J. 2003. *Invasive species: vectors and management strategies*, Island Press.

CATTANEO, A., LUBOWSKI, R., BUSCH, J., CREED, A., STRASSBURG, B., BOLTZ, F. & ASHTON, R. 2010. On international equity in reducing emissions from deforestation. *Environmental Science & Policy,* 13**,** 742-753.

COHEN, A. N. & DOBBS, F. C. 2015. Failure of the public health testing program for ballast water treatment systems. *Marine pollution bulletin,* 91**,** 29-34.

COOPER, W. J., JONES, A. C., WHITEHEAD, R. F. & ZIKA, R. G. 2007. Sunlight-Induced Photochemical Decay of Oxidants in Natural Waters:  Implications in Ballast Water Treatment. *Environmental Science & Technology,* 41**,** 3728-3733.

CORBETT, J. J., WANG, H. & WINEBRAKE, J. J. 2009. The effectiveness and costs of speed reductions on emissions from international shipping. *Transportation Research Part D: Transport and Environment,* 14**,** 593-598.

CORONG, E. L., HERTEL, T. W., MCDOUGALL, R., TSIGAS, M. E. & VAN DER MENSBRUGGHE, D. 2017. The standard GTAP model, version 7. *Journal of Global Economic Analysis,* 2**,** 1-119.

COUNTRYMAN, A. M., FRANCOIS, J. F. & ROJAS-ROMAGOSA, H. 2016. Melting ice caps: implications for Asian trade with North America and Europe. *International Journal of Trade and Global Markets,* 9**,** 325-369.

COUNTRYMAN, A. M., WARZINIACK, T. & GREY, E. 2018. Implications for US trade and nonindigenous species risk resulting from increased economic integration of the Asia-Pacific Region. *Society & Natural Resources,* 31**,** 942-959.

DRAKE, J. M. & LODGE, D. M. 2004. Global hot spots of biological invasions: evaluating options for ballast-water management. *Proceedings of the Royal Society B: Biological Sciences,* 271**,** 575-580.

DRUCKMAN, A. & JACKSON, T. 2008. Measuring resource inequalities: The concepts and methodology for an area-based Gini coefficient. *Ecological Economics,* 65**,** 242-252.

DUMAS, C. F. & WHITEHEAD, J. C. 2008. The Potential Economic Benefits of Coastal Ocean Observing Systems: The Southeast Atlantic Region. *Coastal Management,* 36**,** 146-164.

FERNANDES, J. A., SANTOS, L., VANCE, T., FILEMAN, T., SMITH, D., BISHOP, J. D. D., VIARD, F., QUEIRÓS, A. M., MERINO, G., BUISMAN, E. & AUSTEN, M. C. 2016. Costs and benefits to European shipping of ballast-water and hull-fouling treatment: Impacts of native and non-indigenous species. *Marine Policy,* 64**,** 148-155.

FUGAZZA, M. & HOFFMANN, J. 2017. Liner shipping connectivity as determinant of trade. *Journal of Shipping and Trade,* 2**,** 1.

GLOSTEN ET AL. 2018. Shore-based Ballast Water Treatment in California, Task 10: Cost Analysis. Sacramento, California.

GOODMAN, A. & OLDFIELD, Z. 2004. *Permanent differences? Income and expenditure inequality in the 1990s and 2000s*, IFS Report.

GROOT, L. 2010. Carbon Lorenz curves. *Resource and Energy Economics,* 32**,** 45-64.

GUPTA, V. K., ALI, I., SALEH, T. A., NAYAK, A. & AGARWAL, S. 2012. Chemical treatment technologies for waste-water recycling—an overview. *RSC Advances,* 2**,** 6380-6388.

HEDENUS, F. & AZAR, C. 2005. Estimates of trends in global income and resource inequalities. *Ecological Economics,* 55**,** 351-364.

HERTEL, T., MARTIN, W. & COUNTRYMAN, A. M. 2010. Potential Implications of the Special Safeguard Mechanism (SSM) in Agriculture.

HERTEL, T. W. & DE LIMA, C. Z. 2020. Climate impacts on agriculture: Searching for keys under the streetlight. *Food Policy,* 95**,** 101954.

HERWIG, R. P., CORDELL, J. R., PERRINS, J. C., DINNEL, P. A., GENSEMER, R. W., STUBBLEFIELD, W. A., RUIZ, G. M., KOPP, J. A., HOUSE, M. L. & COOPER, W. J. 2006. Ozone treatment of ballast water on the oil tanker S/T Tonsina: chemistry, biology and toxicity. *Marine Ecology Progress Series,* 324**,** 37-55.

JACOBSON, A., MILMAN, A. D. & KAMMEN, D. M. 2005. Letting the (energy) Gini out of the bottle: Lorenz curves of cumulative electricity consumption and Gini coefficients as metrics of energy distribution and equity. *Energy Policy,* 33**,** 1825-1832.

KALUZA, P., KÖLZSCH, A., GASTNER, M. T. & BLASIUS, B. 2010. The complex network of global cargo ship movements. *Journal of the Royal Society Interface,* 7**,** 1093-1103.

KELLER, R. P., DRAKE, J. M., DREW, M. B. & LODGE, D. M. 2011. Linking environmental conditions and ship movements to estimate invasive species transport across the global shipping network. *Diversity and Distributions,* 17**,** 93-102.

KELLER, R. P., FRANG, K. & LODGE, D. M. 2008. Preventing the spread of invasive species: economic benefits of intervention guided by ecological predictions. *Conservation Biology,* 22**,** 80-88.

KELLER, R. P., LODGE, D. M., LEWIS, M. A. & SHOGREN, J. F. 2009. *Bioeconomics of invasive species: integrating ecology, economics, policy, and management*, Oxford University Press.

KING, D. M., RIGGIO, M. & HAGAN, P. T. 2009. Preliminary Cost Analysis of Ballast Water Treatment Systems.

LOCARNINI, R. A., MISHONOV, A. V., ANTONOV, J. I., BOYER, T. P., GARCIA, H. E., BARANOVA, O. K., ZWENG, M. M., PAVER, C. R., REAGAN, J. R. & JOHNSON, D. R. 2013. World ocean atlas 2013. Volume 1, Temperature.

LORENZ, M. O. 1905. Methods of measuring the concentration of wealth. *Publications of the American statistical association,* 9**,** 209-219.
MILLER, A. W. & RUIZ, G. M. 2014. Arctic shipping and marine invaders. *Nature Climate Change,* 4**,** 413.
MINTON, M. S., VERLING, E., MILLER, A. W. & RUIZ, G. M. 2005. Reducing propagule supply and coastal invasions via ships: effects of emerging strategies. *Frontiers in Ecology and the Environment,* 3**,** 304-308.
NONG, D., WARZINIACK, T., COUNTRYMAN, A. M. & GREY, E. K. 2019. Melting Arctic sea ice: Implications for nonindigenous species (NIS) spread in the United States. *Environmental Science & Policy,* 91**,** 81-91.
PAPATHANASOPOULOU, E. & JACKSON, T. 2009. Measuring fossil resource inequality—A case study for the UK between 1968 and 2000. *Ecological Economics,* 68**,** 1213-1225.
REUSSER, D. A., LEE II, H., FRAZIER, M., RUIZ, G. M., FOFONOFF, P. W., MINTON, M. S. & MILLER, A. W. 2013. Per capita invasion probabilities: an empirical model to predict rates of invasion via ballast water. *Ecological Applications,* 23**,** 321-330.
RUITENBEEK, H. J. 1996. Distribution of ecological entitlements: Implications for economic security and population movement. *Ecological Economics,* 17**,** 49-64.
RUIZ, G. M. & CARLTON, J. T. 2003. Invasion vectors: a conceptual framework for management. *Invasive species: vectors and management strategies***,** 459-504.
RUIZ, G. M. & REID, D. F. 2007. Current state of understanding about the effectiveness of ballast water exchange (BWE) in reducing aquatic nonindigenous species (ANS) introductions to the Great Lakes Basin and Chesapeake Bay, USA: synthesis and analysis of existing information.
SAEBI, M., XU, J., CURASI, S. R., GREY, E. K., CHAWLA, N. V. & LODGE, D. M. 2020a. Network analysis of ballast-mediated species transfer reveals important introduction and dispersal patterns in the Arctic. *Scientific Reports,* 10**,** 19558.
SAEBI, M., XU, J., GREY, E. K., LODGE, D. M., CORBETT, J. J. & CHAWLA, N. 2020b. Higher-order patterns of aquatic species spread through the global shipping network. *Plos one,* 15**,** e0220353.
SALEH, T. A. 2020. Nanomaterials: Classification, properties, and environmental toxicities. *Environmental Technology & Innovation,* 20**,** 101067.
SALEH, T. A. 2021. Protocols for synthesis of nanomaterials, polymers, and green materials as adsorbents for water treatment technologies. *Environmental Technology & Innovation***,** 101821.
SARDAIN, A., SARDAIN, E. & LEUNG, B. 2019. Global forecasts of shipping traffic and biological invasions to 2050. *Nature Sustainability*.
SEEBENS, H., GASTNER, M. T. & BLASIUS, B. 2013. The risk of marine bioinvasion caused by global shipping. *Ecology Letters,* 16**,** 782-790.
SEEBENS, H., SCHWARTZ, N., SCHUPP, P. J. & BLASIUS, B. 2016. Predicting the spread of marine species introduced by global shipping. *Proceedings of the National Academy of Sciences,* 113**,** 5646-5651.
SHACKLETON, R. T., LARSON, B. M. H., NOVOA, A., RICHARDSON, D. M. & KULL, C. A. 2019. The human and social dimensions of invasion science and management. *Journal of Environmental Management,* 229**,** 1-9.
SPALDING, M. D., FOX, H. E., ALLEN, G. R., DAVIDSON, N., FERDAÑA, Z. A., FINLAYSON, M., HALPERN, B. S., JORGE, M. A., LOMBANA, A. & LOURIE, S. A.
19

2007. Marine ecoregions of the world: a bioregionalization of coastal and shelf areas. *BioScience,* 57**,** 573-583.
THE EPA 2012. Economic Impacts of the Category 3 Marine Rule on Great Lakes Shipping.
TOL, R. S., DOWNING, T. E., KUIK, O. J. & SMITH, J. B. 2004. Distributional aspects of climate change impacts. *Global Environmental Change,* 14**,** 259-272.
US ARMY CORPS OF ENGINEERS 2002. Economic Guidance Memorandum 02-06: FY 2002 Deep Draft Vessel Operating Costs. Washington, DC: US Army Corps of Engineers
WANG, Z. & CORBETT, J. J. 2020. Regulation Scenario-based Cost-effectiveness Analysis of Ballast Water Treatment Strategies *Management of Biological Invasions 11 (in press)*.
WANG, Z., NONG, D., COUNTRYMAN, A. M., CORBETT, J. J. & WARZINIACK, T. 2020a. Potential impacts of ballast water regulations on international trade, shipping patterns, and the global economy *Journal of Environmental Management, under revision review*.
WANG, Z., NONG, D., COUNTRYMAN, A. M., CORBETT, J. J. & WARZINIACK, T. 2020b. Potential impacts of ballast water regulations on international trade, shipping patterns, and the global economy: An integrated transportation and economic modeling assessment. *Journal of Environmental Management,* 275**,** 110892.
XU, J., WICKRAMARATHNE, T. L. & CHAWLA, N. V. 2016. Representing higher-order dependencies in networks. *Science advances,* 2**,** e1600028.
ZIMM, C. & NAKICENOVIC, N. 2020. What are the implications of the Paris Agreement for inequality? *Climate Policy,* 20**,** 458-467.
ZWENG, M. M., REAGAN, J. R., ANTONOV, J. I., LOCARNINI, R. A., MISHONOV, A. V., BOYER, T. P., GARCIA, H. E., BARANOVA, O. K., JOHNSON, D. R. & SEIDOV, D. 2013. World ocean atlas 2013. volume 2, salinity.